\documentstyle[preprint,epsf]{jpsj}
\def\runtitle{
Dynamical Properties of 1D 
Multicomponent Quantum Liquids}
\def\runauthor{Satoshi   {\sc Miyashita}, Akira {\sc Kawaguchi} 
and Norio {\sc Kawakami}
} 
\title{Dynamical Properties of One-Dimensional 
Multicomponent Quantum Liquids in Metallic Phase}
\author{Satoshi {\sc  Miyashita}
\footnote{E-mail : satoshi@tp.ap.eng.osaka-u.ac.jp}
, Akira {\sc Kawaguchi}
 and Norio {\sc Kawakami}}
\inst{Department of Applied Physics,
Osaka University, Suita, Osaka 565-0871, Japan}

\recdate{\today}

\abst{%
We investigate low-energy dynamical properties of 
one-dimensional multicomponent quantum liquids 
with the short-range interaction as well as the  
$1/x$-type long-range interaction. 
By calculating the single-particle 
spectrum and the dynamical spin susceptibility by means of 
the bosonization method, 
we discuss how the orbital degeneracy and the band splitting  affect
the dynamical response functions.  The effect of the 
long-range interaction is also addressed.  Although 
the long-range interaction suppresses charge fluctuations,
it effectively enhances  spin fluctuations via the formation of
the Wigner crystal.
}

\kword
{ Tomonaga-Luttinger liquid, dynamical responses, orbital degeneracy, 
long-range interaction}

\begin{document}
\sloppy
\maketitle

%%%%%%%%%%%%%%%%%%%%%%%%%%%%%%%%%%%%%%%%%%%%%%%%%%%%%%%%%%%%%%%%%%%
%%%%%%%%%%%%%%%%%%%%%%%%%%%%%%%%%%%%%%%%%%%%%%%%%%%%%%%%%%%%%%%%%%%
%%%%%%%%%                   　　　　  %%%%%%%%%%%%%%%%%%%%%%%%%%%%%
%%%%%%%%% 1.  Introduction  　　　　  %%%%%%%%%%%%%%%%%%%%%%%%%%%%%
%%%%%%%%%                   　　　　  %%%%%%%%%%%%%%%%%%%%%%%%%%%%%
%%%%%%%%%%%%%%%%%%%%%%%%%%%%%%%%%%%%%%%%%%%%%%%%%%%%%%%%%%%%%%%%%%%
%%%%%%%%%%%%%%%%%%%%%%%%%%%%%%%%%%%%%%%%%%%%%%%%%%%%%%%%%%%%%%%%%%%

\section{Introduction}

Low-energy excitations in one-dimensional (1D) correlated 
electron systems with the short-range interaction are described by 
collective charge and spin fluctuations.  This class 
of quantum liquids is referred to  as the
Tomonaga-Luttinger (TL) liquid.
\cite{Tomonaga,Luttinger,Haldane}
Recently, dynamical properties of 1D electron systems have been
 studied extensively by various experimental methods, 
{\it e.g.} angle resolved photoemission spectroscopy (ARPES)
\cite{Kim,Kobayashi}, 
NMR \cite{Takigawa,Chabo,Goto}, etc.  
In  particular, the spin-charge separation inherent in 1D 
electron systems was observed in the single-particle
spectrum via the ARPES measurements\cite{Kim,Kobayashi}, and 
typical power-law behaviors in  spin correlation functions
were observed in the temperature dependence of
 the NMR relaxation rate \cite{Sachdev,Chitra,Kawaguchi} in various 
1D compounds.\cite{Takigawa,Chabo,Goto}

More recently, the effect of the orbital degrees of freedom has 
attracted renewed interest in correlated electron systems. 
Such orbital effects are indeed important in 1D.
For example, in 1D correlated electron systems at quarter-filling, 
such as Na$_2$Ti$_2$Sb$_2$O\cite{Axtell} and 
NaV$_{2}$O$_{5}$\cite{Isobe,YFujii,Smolinski,Seo,Thalmeier,Mostovoy}, 
the effective orbital degrees of freedom are essential  
to understand the correct low-temperature properties.
In this connection, the 1D multicomponent spin-orbital model 
has been extensively studied analytically
\cite{Sutherland,Affleck,Izergin,Itakura,Azaria,YTsukamoto,Itoi} and 
numerically\cite{Pati,Yamashita}.
However, dynamical correlation functions have not
 been studied systematically so far, 
except for exactly solvable $1/r^{2}$-models\cite{Ha,Kato,Arikawa}.

It is usually assumed that 
 the Coulomb interaction is well screened, and the resulting 
interaction is short-ranged in the TL liquid. However, 
this is not always the 
case for quasi-1D systems, since the screening effect due to
adjacent chains is not efficient enough in some cases, leading 
to the bare long-range interaction.  The resulting 
quantum liquid is a 1D analog of the Wigner
crystal, which exhibits quite different properties from those 
of the TL liquid.\cite{Schulz,Otani,Tsukamoto}

In this paper, we investigate the effect of the
orbital degeneracy on low-energy dynamical properties of the TL
liquid by employing  the bosonization method for  the two-orbital
electron model.  We also
 clarify  how the long-range interaction affects
dynamical properties via the formation of the 1D Wigner crystal.
This paper is organized as follows. 
In \S 2, we introduce the model and derive a low-energy 
effective Hamiltonian in the bosonized form. 
In \S 3, we investigate the orbital effects on  the TL liquid 
with the short-range interaction and then in \S 4
 discuss the effect of  the long-range interaction on
the low-energy properties of the system.
A brief summary is given in the last section. 

%%%%%%%%%%%%%%%%%%%%%%%%%%%%%%%%%%%%%%%%%%%%%%%%%%%%%%%%%%%%%%%%%%%
%%%%%%%%%%%%%%%%%%%%%%%%%%%%%%%%%%%%%%%%%%%%%%%%%%%%%%%%%%%%%%%%%%%
%%%%%%%%%                   　　　　     %%%%%%%%%%%%%%%%%%%%%%%%%%
%%%%%%%%% 2. Model and Formulation       %%%%%%%%%%%%%%%%%%%%%%%%%%
%%%%%%%%%      　　　　                  %%%%%%%%%%%%%%%%%%%%%%%%%%
%%%%%%%%%                   　　　　     %%%%%%%%%%%%%%%%%%%%%%%%%%
%%%%%%%%%%%%%%%%%%%%%%%%%%%%%%%%%%%%%%%%%%%%%%%%%%%%%%%%%%%%%%%%%%%
%%%%%%%%%%%%%%%%%%%%%%%%%%%%%%%%%%%%%%%%%%%%%%%%%%%%%%%%%%%%%%%%%%%

\section{ Low-Energy Effective Hamiltonian}

In this section, we outline the derivation of a 
low-energy effective Hamiltonian in the bosonized form. 
We start with the 1D two-orbital Hubbard model 
with the band splitting ${\Delta}$,
%%%%%%%%%%%%%%%%%%%%%%%%%%%%%%%%%%%%%%%%%%%%%%%%%%%%%%%%%%%%%%%%
\begin{eqnarray}
{\cal H} = &-&t \sum_{j a \sigma} 
               (c_{j a \sigma}^{\dag} c_{j+1 a \sigma} + h.c.) \nonumber\\
           &+& \frac{1}{2} \sum_{i,j} \sum_{a,b} \sum_{\sigma,\sigma'}
                  U(x_{i}-x_{j})
                  n_{i a \sigma} n_{j b \sigma'} 
                  (1-\delta_{i,j}\delta_{a,b}\delta_{\sigma,\sigma'})  
\nonumber\\
         &-& \Delta \sum_{j} {[(n_{j 1 \uparrow}+n_{j 1 \downarrow})
                            -(n_{j 2 \uparrow}+n_{j 2 \downarrow})]},
\label{HB_EQ}
\end{eqnarray}
%%%%%%%%%%%%%%%%%%%%%%%%%%%%%%%%%%%%%%%%%%%%%%%%%%%%%%%%%%%%%%%%
where ${\it c}_{j a \sigma}^{\dag}$ creates an electron 
with the orbital index ${\it a}$ = 1(lower band), 2(upper band) and 
spin ${\sigma}$ = ${\uparrow}, {\downarrow}$ at the $j$-th site. 
In the absence of the band splitting, ${\Delta}$=0, the above 
model corresponds to the  SU(4) symmetric Hubbard model.
For simplicity, we have neglected the Hund-rule coupling
between different orbitals, which are irrelevant in the 
following discussions in the TL liquid phase.

Passing to the continuum limit, we derive a
low-energy effective Hamiltonian in terms of the 
slowly varying right- and left-going Fermi fields
 $c^{(+)}_{j a \sigma}$ and $c^{(-)}_{j a \sigma}$, which
 are related to the original lattice operator as
%%%%%%%%%%%%%%%%%%%%%%%%%%%%%%%%%%%%%%%%%%%%%%%%%%%%%%%%%%%%%%%%%%%%%%%%%%
\begin{equation}
  c_{j a \sigma}
   \sim \sqrt{a_{0}} 
       \left[
        {\rm e}^{{\rm i}k_{F}^{(a)}ja_{0}}c^{(+)}_{j a \sigma}
       +{\rm e}^{-{\rm i}k_{F}^{(a)}ja_{0}}c^{(-)}_{j a \sigma}
       \right],
\label{eq:c+-}
\end{equation}
%%%%%%%%%%%%%%%%%%%%%%%%%%%%%%%%%%%%%%%%%%%%%%%%%%%%%%%%%%%%%%%%%%%%%%%%%%
where $a_{0}$ is the lattice spacing and $k_{F}^{(1)}$ 
($k_{F}^{(2)}$) is the Fermi momentum of the lower (upper) band. 
Since we are interested in a metallic
phase with the repulsive interaction, we 
are left with only forward scatterings. 
By using the Fourier decompositions 
$c^{(r)}_{ja\sigma}=\frac{1}{\sqrt{N}}\sum_{p}{\rm e}^{-{\rm i}pja_{0}}
c_{rpa\sigma}$ ($r$=+ or ${\scriptstyle -}$), 
 the Hamiltonian ($\ref{HB_EQ}$) is rewritten as 
%%%%%%%%%%%%%%%%%%%%%%%%%%%%%%%%%%%%%%%%%%%%%%%%%%%%%%%%%%%%%%%%%%%%%%%%%%
\begin{eqnarray}
{\cal H} &=&  \frac{\pi}{L} \sum_{p,a,\sigma} {\it v}^{(a)}_{F} 
              [
                \rho_{+ a \sigma}(p) \rho_{+ a \sigma}(-p)
              + \rho_{- a \sigma}(-p) \rho_{- a \sigma}(p)
              ]
\nonumber \\
         &+&  \frac{1}{2 L} 
              \sum_{\stackrel{\scriptstyle p,a,b}{\sigma,\sigma'}}
              U(p) 
              [
                \rho_{+ a \sigma}(p) + \rho_{- a \sigma}(p)
              ]
              [ 
                \rho_{+ b \sigma'}(-p) + \rho_{- b \sigma'}(-p)
              ],
\nonumber \\
\label{TL_eq}
\end{eqnarray}
%%%%%%%%%%%%%%%%%%%%%%%%%%%%%%%%%%%%%%%%%%%%%%%%%%%%%%%%%%%%%%%%%%%%%%%%%%
where $\rho_{ra\sigma}(p)=
\sum_{k}c^{\dag}_{rk+pa\sigma}c_{rka\sigma}$ are the
density operators satisfying the bosonic commutation relation,
and  ${\it v}^{(a)}_{F}=a_{0}t\sin(k^{(a)}_{F}a_{0})$ is the Fermi velocity 
of the lower band ($a=1$) or the upper band ($a=2$). 
We have four kinds of bosonic fields, $\rho_{r1\uparrow}$, 
$\rho_{r1\downarrow}$, $\rho_{r2\uparrow}$ and $\rho_{r2\downarrow}$, 
which are converted via a unitary transformation
to the new bosonic fields corresponding to the 'charge'(c), 'spin'(s), 
'orbital'(o) and 'spin-orbital'(so) sectors,
%%%%%%%%%%%%%%%%%%%%%%%%%%%%%%%%%%%%%%%%%%%%%%%%%%%%%%%%%%%%%%%%%%%%%%%%%%
\begin{eqnarray}
\left(
\begin{array}{c}
\rho_{r c}(p) \\
\rho_{r s}(p) \\
\rho_{r o}(p) \\
\rho_{r so}(p) \\
\end{array}
\right)
= \frac{1}{2}
\left(
\begin{array}{cccc}
1 & 1 & 1 & 1 \\
1 & -1 & 1 & -1 \\
1 & 1 & -1 & -1 \\
1 & -1 & -1 & 1 \\
\end{array}
\right)
\left(
\begin{array}{c}
\rho_{r 1 \uparrow}(p) \\
\rho_{r 1 \downarrow}(p) \\
\rho_{r 2 \uparrow}(p) \\
\rho_{r 2 \downarrow}(p) \\
\end{array}
\right).
\nonumber \\
\end{eqnarray}
%%%%%%%%%%%%%%%%%%%%%%%%%%%%%%%%%%%%%%%%%%%%%%%%%%%%%%%%%%%%%%%%%%%%%%%%%%
Moreover, when the band splitting takes a finite value, it is convenient to
use the new basis for the charge and orbital sectors, 
%%%%%%%%%%%%%%%%%%%%%%%%%%%%%%%%%%%%%%%%%%%%%%%%%%%%%%%%%%%%%%%%%%%%%%
\begin{eqnarray}
%&&
\left(
\begin{array}{c}
\rho_{+ c} + \rho_{- c} \\
\rho_{+ o} + \rho_{- o}\\
\end{array}
\right)
%\nonumber
%\\
%&&
=
\left(
\begin{array}{cc}
\xi_1(p) & -\xi_3(p) \\
\xi_2(p) & \xi_1(p) \\
\end{array}
\right)
\left(
\begin{array}{c}
\tilde{\rho}_{+ c} + \tilde{\rho}_{- c} \\
\tilde{\rho}_{+ o} + \tilde{\rho}_{- o} \\
\end{array}
\right),
%\nonumber \\
\label{linear_trans}
\end{eqnarray}
%%%%%%%%%%%%%%%%%%%%%%%%%%%%%%%%%%%%%%%%%%%%%%%%%%%%%%%%%%%
where
%%%%%%%%%%%%%%%%%%%%%%%%%%%%%%%%%%%%%%%%%%%%%%%%%%%%%%%%%%
\begin{eqnarray}
\xi_1(p)&=&\cos[\alpha(p)],  \\
\xi_2(p)&=&y(p) \sin[\alpha(p)], \\
\xi_3(p)&=&\frac{1}{y(p)}\sin[\alpha(p)]. 
\end{eqnarray}
%%%%%%%%%%%%%%%%%%%%%%%%%%%%%%%%%%%%%%%%%%%%%%%%%%%%%%%%%%%
%%Here $\alpha(p)$ is the rotation angle in the charge-orbital space. 
The explicit formulae for $y(p)$ and $\alpha(p)$ are given in Appendix. 
Note that 
when two orbitals are degenerate ($\Delta$=0), 
we have  $\alpha(p)=0$, so that $\xi_1=1$ and $\xi_2 =\xi_3=0$.

Consequently, we end up with
the Hamiltonian, 
%%%%%%%%%%%%%%%%%%%%%%%%%%%%%%%%%%%%%%%%%%%%%%%%%%%%%%%%%%%%
\begin{eqnarray}
{\cal H} 
%&=& 
= \frac{\pi}{2L} \sum_{\nu} \sum_{p}
        \tilde{{\it v}}_{\nu}(p) \left\{
           \frac{1}{\tilde{K}_{\nu}(p)} 
            [\tilde{\rho}_{+ \nu}(p)+\tilde{\rho}_{- \nu}(p)]^{2}
                                 \right.
%
%\nonumber \\
%         &&\ \ \ \ \ \ \ \ \ \ \ \ \ \ \ \ \ \ \ \ 
            \left.
                 +  \tilde{K}_{\nu}(p) 
             [\tilde{\rho}_{+ \nu}(p)-\tilde{\rho}_{- \nu}(p)]^{2}
                        \right\},
\label{L-TL_eq}
\end{eqnarray}
%%%%%%%%%%%%%%%%%%%%%%%%%%%%%%%%%%%%%%%%%%%%%%%%%%%%%%%%%%%%%%%%%%%%%%%%%%
where the velocities and the TL parameters are summarized in 
Appendix.

This completes the derivation of a low-energy effective
theory of 1D correlated electrons with two orbitals.
Though we have outlined the way in terms of the Hubbard model,
the resulting formula  can be applied to 
generic two-band models in a metallic phase.

Before closing this section, some comments are in order for the 
formula (\ref{L-TL_eq}).
In terms of the spin and spin-orbital basis, the velocities
read $\tilde{{\it v}}_{s}$=${\it v}_{F}^{(1)}$ and 
$\tilde{{\it v}}_{so}$=$\tilde{{\it v}}_{F}^{(2)}$. 
Moreover, we should set $\tilde{K}_{s}$=$\tilde{K}_{so}$=1 
according to SU(2) symmetry required for these excitation modes. 
%%As for the charge and orbital modes, we rescale the parameters as 
%%$\tilde{{\it v}}_{c,f}(p)$$\rightarrow$
%%$\frac{\tilde{{\it v}}_{c,f}}
%%{\tilde{{\it v}}_{c,f}(\infty)}$$\tilde{{\it v}}_{c,f}(p)$ and 
%%$\tilde{K}_{c,f}(p)$$\rightarrow$
%%$\frac{\tilde{K}_{c,f}}
%%{\tilde{K}_{c,f}(\infty)}$$\tilde{K}_{c,f}(p)$. 
%

%%%%%%%%%%%%%%%%%%%%%%%%%%%%%%%%%%%%%%%%%%%%%%%%%%%%%%%%%%%%%%%%%%%%%%%%%%
%%%%%%%%%%%%%%%%%%%%%%%%%%%%%%%%%%%%%%%%%%%%%%%%%%%%%%%%%%%%%%%%%%%
%%%%%%%%%%%%%%%%%%%%%%%%%%%%%%%%%%%%%%%%%%%%%%%%%%%%%%%%%%%%%%%%%%%
%%%%%%%%%                   　　　　      %%%%%%%%%%%%%%%%%%%%%%%%%
%%%%%%%%% 2. The  short-range interacting %%%%%%%%%%%%%%%%%%%%%%%%%
%%%%%%%%%    case         　　　　        %%%%%%%%%%%%%%%%%%%%%%%%%
%%%%%%%%%                   　　　　      %%%%%%%%%%%%%%%%%%%%%%%%%
%%%%%%%%%%%%%%%%%%%%%%%%%%%%%%%%%%%%%%%%%%%%%%%%%%%%%%%%%%%%%%%%%%%
%%%%%%%%%%%%%%%%%%%%%%%%%%%%%%%%%%%%%%%%%%%%%%%%%%%%%%%%%%%%%%%%%%%

\section{Multicomponent Tomonaga-Luttinger Liquid}

We first consider the system with the short-range interaction and 
discuss how the orbital degrees of freedom affect the 
dynamical response functions. 
When the interaction is short-ranged ($\delta$-function), 
we set all the parameters, $\tilde{{\it v}}_{\nu}(p)$ 
and $\tilde{K}_{\nu}(p)$, 
independent of the momentum $p$.  We employ the {\it t-J} 
model as a  microscopic model
 to give the values of  $\tilde{{\it v}}_{c,o}^{'}$ 
and $\tilde{K}_{c,o}^{'}$.
Furthermore, if we make use of the integrable supersymmetric
{\it t-J} model\cite{Sutherland,Schlottmann,Bares,Kawakami}, 
where the hopping $t$ and the spin
exchange coupling $J$ satisfies the relation, $t=J$,
the above TL parameters can be determined by the Bethe ansatz method 
combined with conformal field theory. 
This model is somewhat special, but 
it can well describe general properties of 
the doped spin-orbital model, as far as we are 
concerned with a metallic system near the insulating phase.

We first derive the single-particle Green function
\cite{Ogata,Sorella,Meden,Nakamura} 
for the TL liquid 
with orbital degeneracy, which is defined as
%%%%%%%%%%%%%%%%%%%%%%%%%%%%%%%%%%%%%%%%%%%%%%%%%%%%%%%%%%%%%%%%%%%%%%%%%%%%%%%
\begin{eqnarray}
\tilde{G}_{r a \sigma}(x,t) 
    &=& 
       - {\rm i} \Theta(t) 
          \langle 
           [ \psi_{r a \sigma}(x,t), \psi_{r a \sigma}^{\dag}(0,0) ]_{+} 
          \rangle
\nonumber \\
    &=&
       - {\rm i} \Theta(t) 
	  [G_{r a \sigma}^{<}(x,t) + G_{r a \sigma}^{<}(-x,-t)],
\label{tilde_GF}
\end{eqnarray}
%%%%%%%%%%%%%%%%%%%%%%%%%%%%%%%%%%%%%%%%%%%%%%%%%%%%%%%%%%%%%%%%%%%%%%%%%%%%%%%
where
%%%%%%%%%%%%%%%%%%%%%%%%%%%%%%%%%%%%%%%%%%
\begin{eqnarray}
G_{r a \sigma}^{<}(x,t) =
     < \psi_{r a \sigma}(x,t) \psi_{r a \sigma}^{\dag}(0,0) >,
\end{eqnarray}
%%%%%%%%%%%%%%%%%%%%%%%%%%%%%%%%%
and $G_{r a \sigma}^{>}(x,t)$ is defined similarly.
The Fermi field is written down as 
%%%%%%%%%%%%%%%%%%%%%%%%%%%%%%%%%%%%%%%%%%%%%%%%%%%%%%%%%%%%%%%%%%%%%%%%%%
\begin{eqnarray}
\psi_{r a \sigma} =                
               \lim_{a_{0} \rightarrow 0}
               \frac{\eta_{r a \sigma}}{\sqrt{2 \pi a_{0}}} 
                 {\rm exp} \left(
                  - \frac{{\rm i}}{2} \sum_{\nu}
                       \left[
                           r \tilde{\phi}_{\nu}
                         - \tilde{\theta}_{\nu}
                       \right]
		           \right),
\label{eq:Fermi3}
\end{eqnarray}
%%%%%%%%%%%%%%%%%%%%%%%%%%%%%%%%%%%%%%%%%%%%%%%%%%%%%%%%%%%%%%%%%%%%%%%%%%
where $\eta_{r a \sigma}$ is the  Klein factor,  which 
ensures the anti-commutation relation between 
the right- and left-going electrons. The above two phase fields are 
%%%%%%%%%%%%%%%%%%%%%%%%%%%%%%%%%%%%%%%%%%%%%%%%%%%%%%%%%%%%%%%%%%%%%%%%%%
\begin{eqnarray}
\tilde{\phi}_{\nu}(x) &=& - \frac{{\rm i} \pi}{L} \sum_{p} \frac{1}{p}
                  {\rm e}^{-\frac{1}{2} a_{0} |p| - {\rm i} p x}
                \xi_{\phi a \sigma}^{\nu}(p)
                [\tilde{\rho}_{+ \nu}(p)+\tilde{\rho}_{- \nu}(p)],
\nonumber \\
\\
\tilde{\theta}_{\nu}(x) &=& \frac{{\rm i} \pi}{L} \sum_{p} \frac{1}{p}
                  {\rm e}^{-\frac{1}{2} a_{0} |p| - {\rm i} p x}
                \xi_{\theta a \sigma}^{\nu}(p)
                [\tilde{\rho}_{+ \nu}(p)-\tilde{\rho}_{- \nu}(p)],
\nonumber \\
\label{Bosonic_Fields}
\end{eqnarray}
%%%%%%%%%%%%%%%%%%%%%%%%%%%%%%%%%%%%%%%%%%%%%%%%%%%%%%%%%%%%%%%%%%%%%%%%%%
where 
%%%%%%%%%%%%%%%%%%%%%%%%%%%%%%%%%%%%%%%%%%%%%%%%%%%%%%%%%%%%%%%%%%%%%%%%%%
\begin{eqnarray}
&&
\left\{
\begin{array}{l}
\xi_{\phi a \sigma}^{c}(p) = 
              \xi_1(p) + (-1)^{a+1} \xi_2(p) 
\\
\xi_{\theta a \sigma}^{c}(p) = 
          \xi_1(p) + (-1)^{a+1} \xi_3(p)
\end{array},
\right.
\\
&&
%\end{eqnarray}
%\begin{eqnarray}
\left\{
\begin{array}{l}
\xi_{\phi a \sigma}^{o}(p) =
           - \xi_3(p) + (-1)^{a+1}\xi_1(p)
\\
\xi_{\theta a \sigma}^{o}(p) =
           - \xi_2(p) + (-1)^{a+1} \xi_1(p)
\end{array},
\right.
\\
&&
%\end{eqnarray}
%\begin{eqnarray}
\left\{
\begin{array}{l}
\xi_{\phi a \sigma}^{s}(p) = \xi_{\theta a \sigma}^{s}(p) = 
\frac{\sigma}{\sqrt{2}} \{1 + (-1)^{a+1} \} 
\\
\xi_{\phi a \sigma}^{so}(p) = \xi_{\theta a \sigma}^{so}(p) =
\frac{\sigma}{\sqrt{2}} \{1 - (-1)^{a+1} \} 
\end{array}.
\right.
\end{eqnarray}
%%%%%%%%%%%%%%%%%%%%%%%%%%%%%%%%%%%%%%%%%%%%%%%%%%%%%%%%%%%%%%%%%%%%%%%%%%
Note that $\tilde{\theta}_{\nu}$ is 
the bosonic field conjugate to $\tilde{\phi}_{\nu}$,
which satisfies 
$[\tilde{\phi}_{\nu}(x) , \tilde{\theta}_{\nu'}(x')]
={\rm i}\frac{\pi}{2}\delta_{\nu,\nu'}{\rm sgn}(x-x')$.
%%%%%%%%%%%%%%%%%%%%%%%%%%%%%%%%%%%%%%%%%%%%%%%%%%%%%%%%%%%%%%%%%%%%%%%%%%

Inserting these expressions into the Green function,
we have
%%%%%%%%%%%%%%%%%%%%%%%%%%%%%%%%%%%%%%%%%%%%%%%%%%%%%%%%%%%%%%%%%%%%%%%%%%%%%%%
\begin{eqnarray}
% && 
G_{r a \sigma}^{<}(x,t)
%\nonumber \\
%   &=& 
 =
    \frac{1}{2 \pi}
    \prod_{\stackrel{\scriptstyle \nu=c,s,}{o,so}}
\frac{{\rm e}^{-{\rm i} \frac{\pi}{2} 
                 2 \tilde{\Delta}_{\nu}^{+} 
                 {\rm sgn}(x+\tilde{{\it v}}_{\nu}t)}
      {\rm e}^{{\rm i} \frac{\pi}{2} 
                 2 \tilde{\Delta}_{\nu}^{-} 
                 {\rm sgn}(x-\tilde{{\it v}}_{\nu}t)}
        \left(
            \frac{\pi}{\beta \tilde{{\it v}}_{\nu}}
        \right)^{2 \tilde{x}_{\nu}}}
           { \left|
              \sinh     
              \left[       
                    \frac{\pi}{\beta \tilde{{\it v}}_{\nu}}
                    (x+\tilde{{\it v}}_{\nu}t)
              \right]
             \right|^{2\tilde{\Delta}_{\nu}^{+}}
             \left|
              \sinh     
              \left[       
                    \frac{\pi}{\beta \tilde{{\it v}}_{\nu}}
                    (x-\tilde{{\it v}}_{\nu}t)
              \right]
             \right|^{ 2\tilde{\Delta}_{\nu}^{-} } },
\nonumber \\
\label{GF}
\end{eqnarray}
%%%%%%%%%%%%%%%%%%%%%%%%%%%%%%%%%%%%%%%%%%%%%%%%%%%%%%%%%%%%%%%%%%%%%%%%%%%%%%%
with $\beta=1/k_{B}T$, where the conformal dimensions are given by
%%%%%%%%%%%%%%%%%%%%%%%%%%%%%%%%%
\begin{eqnarray}
2\tilde{\Delta}_{\nu}^{\pm}
=\frac{1}{16} [\tilde{K}_{\nu}
{\xi_{\phi a \sigma}^{\nu}}^{2}+
{\xi_{\theta a \sigma}^{\nu}}^{2}/\tilde{K}_{\nu}
{\mp} 2 r
\xi_{\phi a \sigma}^{\nu}\xi_{\theta a \sigma}^{\nu}], 
\end{eqnarray}
%%%%%%%%%%%%%%%%%%%%%%%%%%%%%%%%
and the corresponding scaling dimensions are
$\tilde{x}_{\nu}$=$\tilde{\Delta}_{\nu}^{+}$+$\tilde{\Delta}_{\nu}^{-}$. 
For $\Delta$=0, we have
$\xi_{\phi a \sigma}^{\nu}$=$\xi_{\theta a \sigma}^{\nu}$ =1,
which gives the conformal dimensions for the SU(4) spin model.\cite{Affleck}
Equation (\ref{GF}) implies that the spin-charge-orbital
 separation occurs in elementary excitations,
as should be expected  for 1D electron systems. 
 Note, however, that 
all these degrees of freedom are combined together to contribute to 
the single-particle Green function.
Fourier transformation  of (\ref{tilde_GF}) 
gives the single-particle spectrum,  $G_{r a \sigma}(k,\omega)$.
%%%%%%%%%%%%%%%%%%%%%%%%%%%%%%%%%%%%%%%%%%%%%%%%%%%%%%%%%%%%%%%%%%%%%%%%%%%%%%%
%%\begin{eqnarray}
%%  G_{r a \sigma}(k,\omega) &=& 
%%                  \int_{- \infty}^{\infty} dt
%%                  \int_{- \infty}^{\infty} dx
%%                   {\rm e}^{{\it i} (\omega t - k x)} 
%%                    \tilde{G}_{r a \sigma}(x,t)
%%\label{FT_GF}
%%\end{eqnarray}
%%%%%%%%%%%%%%%%%%%%%%%%%%%%%%%%%%%%%%%%%%%%%%%%%%%%%%%%%%%%%%%%%%%%%%%%%%%%%%%

Let us now investigate the single-particle spectrum by
focusing on right-moving fermions ($r$=+). 
To see the effect of the orbital degeneracy, 
we first compare the spectral function of the 
single-orbital electron system (2-component
TL liquid) with that of the two-orbital one
(4-component TL liquid). 
The system has the dispersions, $\omega$=$\tilde{\it v}_{\nu}$$k$, 
so that the  spectral function may have several peaks 
around $\omega$=$\tilde{\it v}_{\nu}$$k$ at finite temperatures.
We use the parameters which are derived 
from the exact solution of the multicomponent supersymmetric 
{\it t-J} model \cite{Sutherland,Itakura} 
near the Mott insulator in order to
parameterize the velocities and the TL parameters 
as a function of the band splitting.  
For the case of single- and two-orbital TL liquids, 
we set $\tilde{{\it v}}_{s}$=2$\pi/n$, 
$\tilde{K}_{c}=1/n $ ($n=2$ or 4) 
whereas $\tilde{K}_{s}$=1 is fixed due to SU(2) or SU(4)
symmetry. We set a small value for the charge velocity 
($\tilde{{\it v}}_{c}$=0.1) 
because we are interested in the vicinity of the
Mott insulator. 
Note that we will measure the energy in the unit of 
the spin exchange coupling $J$.
%%%%%%%%%%%%%%%%%%%%%%%%%%%%%%%%%%%%%%%%%%%%%%%%%%%%%%%%%%%%%%%%%%%%%%%%%%%%
\begin{figure}[thb]
\begin{center}
%\vspace{-0.cm}
%\hspace{0.0cm}
\leavevmode \epsfxsize=150mm 
\epsffile{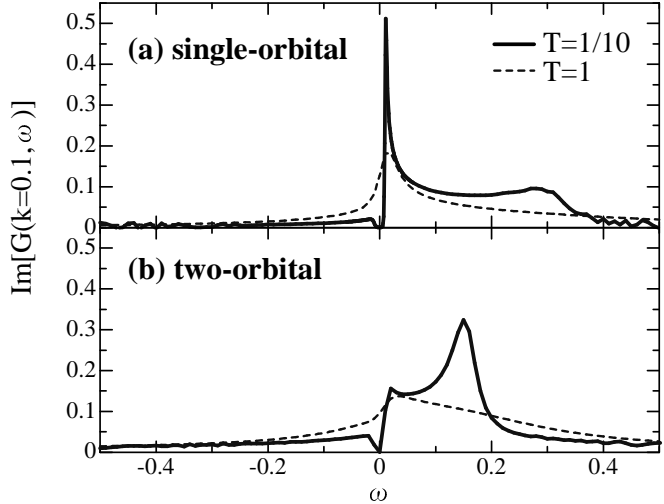}
%\vspace{-2.2cm}
\end{center}
\caption{Single-particle spectrum for the TL liquid; 
(a) SU(2) single-orbital case, 
$\tilde{{\it v}}_{c}$=0.1, $\tilde{{\it v}}_{s}$=$\pi$,
(b) SU(4) two-orbital case, $\tilde{{\it v}}_{c}$=0.1, 
$\tilde{{\it v}}_{s}$=$\tilde{{\it v}}_{o}$=
$\tilde{{\it v}}_{so}$=$\pi$/2.
}
\label{2+4comgf}
\end{figure}
%%%%%%%%%%%%%%%%%%%%%%%%%%%%%%%%%%%%%%%%%%%%%%%%%%%%%%%%%%%%%%%%%%%

We show the results for the TL liquid with and without orbital degeneracy 
in Fig. $\ref{2+4comgf}$.  At high temperatures, all the kinds of
excitations show over-damped behaviors in which the spectral weight
is distributed broadly around $\omega \sim 0$.  This is 
valid both for the single- and two-orbital models. 
As the temperature decreases, the peak structure of 
the spectrum is  
gradually developed around each mode of elementary 
excitations.  In the single-orbital case, 
the peak around the charge mode ($\omega$$\sim$0.01) 
is enhanced more prominently
 than that for the spin mode ($\omega$$\sim$0.3)
at low temperatures.   On the other hand, 
in the two-orbital case,  the peak structure for the charge mode
 ($\omega$$\sim$0.01) is rather suppressed, 
while that for the spin mode ($\omega$$\sim$0.17) is enhanced. 
This is understood  by looking at the zero-temperature 
behavior of the spectrum. At $T=0$, the above spectral function
shows power-law divergence near each excitation mode as,
%%%%%%%%%%%%%%%%%%%%%%%%%%%%%%%%%
\begin{eqnarray}
{\rm Im} G(k, \omega) 
\sim |\omega - \tilde{v}_\nu k|^{-2\tilde{\Delta}_\nu^-}, 
\hskip 4mm
\omega >0,
\end{eqnarray}
%%%%%%%%%%%%%%%%%%%%%%%%%%%%%%%%
for the mode specified by $\nu$,
 where the corresponding critical exponents are listed in Table
\ref{tbl:exponent1}.  It is seen that 
the spin exponent $2\tilde{\Delta}_s^-$ is increased in the two-orbital case
due to the orbital degeneracy, resulting in the
enhanced peak structure.  On the other hand, the charge exponent 
$2\tilde{\Delta}_c^-$ is decreased in the orbitally degenerate case, 
being consistent 
with the suppressed charge mode in Fig. $\ref{2+4comgf}$(b).
 In this way, the orbital degeneracy 
affects not only the internal spin- and orbital-degrees of 
freedom but also  the charge degrees of freedom, changing 
the structure of the spectral function rather considerably. 

\begin{figure}[thb]
\begin{center}
%\vspace{-1.cm}
%\hspace{+3.0cm}
\leavevmode \epsfxsize=150mm 
\epsffile{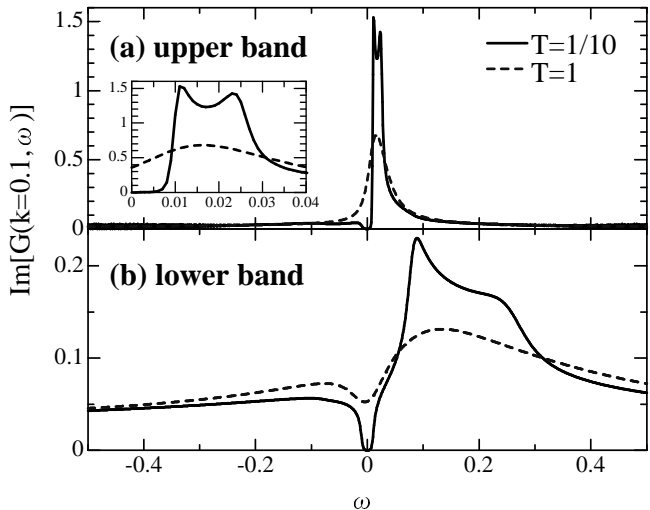}
%\vspace{-2.2cm}
%\hspace{-5.0cm}
\end{center}
\caption{Single-particle spectrum for the 
two-orbital TL liquid 
with the band splitting $\Delta$$\simeq$0.6; 
$\tilde{{\it v}}_{c}$=0.1, $\tilde{{\it v}}_{s}$$\simeq$2.6, 
$\tilde{{\it v}}_{o}$$\simeq$0.8, $\tilde{{\it v}}_{so}$$\simeq$0.25, 
(a)upper band, (b)lower band.}
\label{4comgf+o}
\end{figure}

We now discuss the effect of the band splitting. 
The single-particle spectral function is shown for 
the band splitting $\Delta$$\simeq$0.6 in Fig. $\ref{4comgf+o}$,
which should be compared with the case of $\Delta$=0 in 
 Fig. $\ref{2+4comgf}$(b).
When the band splitting is present, the velocities 
take different values.   We set these parameters as,
$\tilde{{\it v}}_{s}$$\simeq$$2.6$, 
$\tilde{{\it v}}_{o}$$\simeq$$0.8$, 
$\tilde{{\it v}}_{so}$$\simeq$$0.25$, 
$\tilde{K}_{c}=1/n$, 
$\tilde{K}_{o}$=2/($n\times$0.436) ($n=4$) according to the exact solution 
of the supersymmetric {\it t-J} model near the Mott insulator
\cite{Itakura,Kawaguchi}
and $\tilde{K}_{s}$=$\tilde{K}_{so}$=1 due to the symmetry requirement. 
%
%%TABLE %%%%%%%%%%%%%%%%%%%%%%%%%%
%%%%%%%%%%%%%%%%%%%%%%%%%%%%%%%%%%%%%%%%%%%%%%%%%%%%%%%%%%%%%%%%%%%
\begin{table}
\begin{center}
\vspace{5mm}
\caption{Critical exponent of the single-particle spectrum.}
\begin{tabular*}{85mm}{ccccc} \hline
 & \makebox[8mm]{2$\tilde{\Delta}_{c}^{-}$}
 & \makebox[8mm]{$2\tilde{\Delta}_{s}^{-}$} 
 & \makebox[8mm]{2$\tilde{\Delta}_{o}^{-}$}
 & \makebox[8mm]{$2\tilde{\Delta}_{so}^{-}$}
  \\ \hline
single-orbital & 
$\frac{9}{16}$ & $\frac{1}{2}$ & - & - \\ \hline
two-orbital ($\Delta$=0)&
$\frac{25}{64}$ & $\frac{3}{4}$ & - & - \\ \hline
two-orbital ($\Delta$$\simeq$0.6) & 
 & & & \\
upper band & 
0.563 & 0 & 0.072 & $\frac{1}{2}$
 \\
lower band  & 
0.250 & $\frac{1}{2}$ & 0.540 & 0 \\ 
\hline
\end{tabular*}
\end{center}
\label{tbl:exponent1}
\end{table}
%
%
%%%%%%%%%%%%%%%%%%%%%%%%%%%%%%%%%%%%%%%%%%%%%%%%%%%%%%%%%%%%%%%%%%%%%%%%%%%
%
%
%%%%%%%%%%%%%%%%%%%%%%%%%%%%%%%%%%%%%%%%%%%%%%%%%%%%%%%%%%%%%%%%%%%%%%%%%%%%
%%%%%%%%%%%%%%%%%%%%%%%%%%%%%%%%%%%%%%%%%%%%%%%%%%%%%%%%%%%%%%%%%%%
In order to clarify the properties of the spectrum, we glance at 
the critical exponents again at zero temperature 
listed in Table \ref{tbl:exponent1}.  It is seen in Fig. \ref{4comgf+o}
that the introduction of the band splitting alters the shape of the 
spectrum rather dramatically.  Firstly, we notice that
the peak structure of the charge sector ($\omega$$\sim$0.01) 
is suppressed (enhanced) for the lower 
(upper) band in the presence of the band 
splitting. This is indeed  seen  from the 
change in the critical exponents; $2\tilde{\Delta}_c^-$ 
decreases (increases) for the lower (upper) band with 
the increase of the band splitting, which has a tendency to
smear (pronounce) the singularity around  $\omega =\tilde{\it v}_c k$ at $T=0$.
The remaining two peaks in the spectrum  of
the lower band in Fig. $\ref{4comgf+o}$(b)
 are respectively identified with the contribution from the 
orbital ($\omega$$\sim$0.08) and spin ($\omega$$\sim$0.26)
degrees of freedom.  Note that the spin-orbital degrees of freedom
do not contribute to the spectrum of the lower band.
On the other hand, the spectrum shows a quite different 
behavior for the upper band (Fig. $\ref{4comgf+o}$(a)). 
Namely, it has two peaks, which are mainly contributed by the charge
and the spin-orbital ($\omega$$\sim$0.025) sectors. As seen from the 
Table \ref{tbl:exponent1}, the orbital sector
also  has a finite contribution, but its critical exponent
is too small to make a visible  peak structure around  $T=1/10$. 
In this way, the introduction of the band splitting 
brings about a considerable change in the profile of the 
spectral function, which is quite contrasted to that expected for
the Fermi liquid system. 
%In particular, it is necessary to 
%carefully identify  which sector of the excitations really
%contributes to the spectral function,
%making a characteristic structure.

\begin{figure}[thb]
\begin{center}
%\vspace{-1.cm}
%\hspace{+3.0cm}
\leavevmode \epsfxsize=150mm 
\epsffile{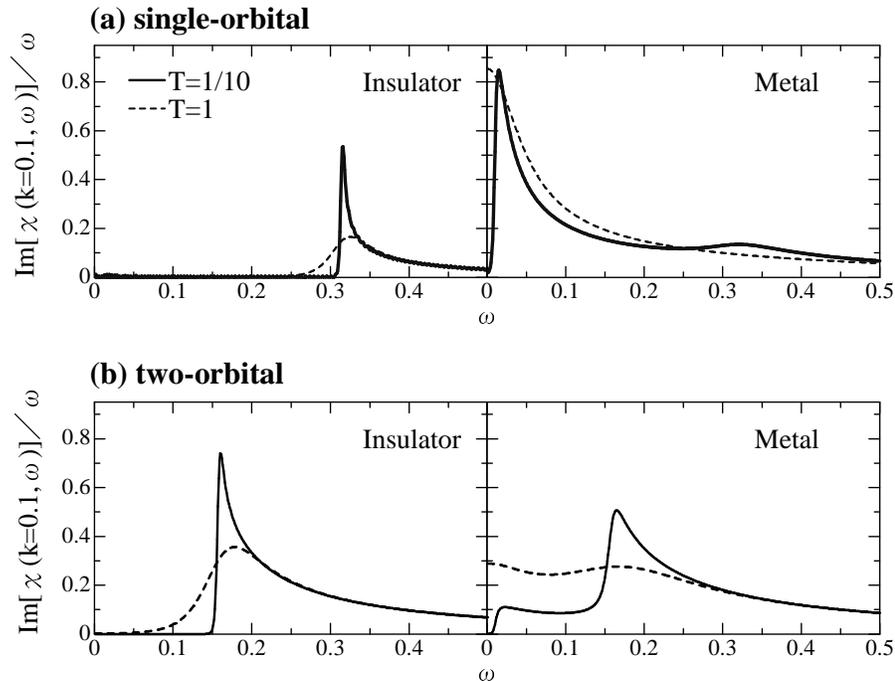}
%\vspace{-0.7cm}
%\hspace{-5.0cm}
\end{center}
\caption{
Dynamical spin susceptibility at (left) and near (right)
 the Mott insulator; 
(a) SU(2) single-orbital case, 
$\tilde{{\it v}}_{c}$=0.1, $\tilde{{\it v}}_{s}$=$\pi$, 
(b) SU(4) two-orbital case, $\tilde{{\it v}}_{c}$=0.1, 
$\tilde{{\it v}}_{s}$=$\tilde{{\it v}}_{o}$=
$\tilde{{\it v}}_{so}$=$\pi$/2. 
The momentum $k$ measures the difference from $\pi$/2. 
}
\label{2+4comxi}
\end{figure}

\begin{figure}[thb]
\begin{center}
%\vspace{0.5cm}
%\hspace{+3.0cm}
\leavevmode \epsfxsize=150mm 
\epsffile{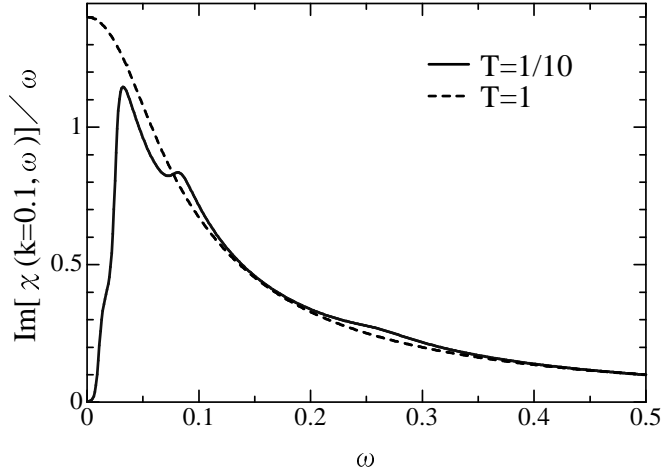}
%\vspace{-2.6cm}
%\hspace{-5.0cm}
\end{center}
\caption{Dynamical spin susceptibility for the 
two-orbital TL liquid in a metallic phase
with the band  splitting $\Delta\simeq$0.6; 
$\tilde{{\it v}}_{c}$=0.1, 
$\tilde{{\it v}}_{s}$$\simeq$2.6, $\tilde{{\it v}}_{o}$=$\simeq$0.8, 
$\tilde{{\it v}}_{so}$$\simeq$0.25. }
\label{4comxi+o}
\end{figure}

We now turn to the dynamical spin susceptibility. 
We show the results calculated for  Im[$\chi$($k$,$\omega$)]/$\omega$
at and near the Mott insulator in Fig. \ref{2+4comxi}.  We also compare
 the results for the single- and two-orbital 
 TL liquid ($\Delta$=0). As is clearly seen from  
the single-orbital case, a considerable amount of the spectral weight 
is  transferred from the spin part to the charge part 
upon hole doping.  This dramatic change is 
in contrast to our naive expectation that the charge excitation 
has little effect on the dynamical spin susceptibility.
It is seen from Fig. \ref{2+4comxi}(b)
that this tendency is also the case for the 
two-orbital model, although the rate of the 
transferred weight is somehow small.
These remarkable results imply that the NMR relaxation rate
\cite{Sachdev,Chitra,Kawaguchi,Takigawa,Chabo,Goto} 
may be possibly enhanced upon hole doping 
via a relaxation process mediated indirectly by
 low-energy charge channels. In Fig. \ref{4comxi+o}
the effect of  the band splitting is shown ($\Delta$$\simeq$0.6).
In this case, the spectral weight is transferred to 
low-energy charge and orbital modes upon hole doping.

We have so far seen that hole-doping induces the considerable
transfer of the spectral weight to the low-energy charge mode
in the dynamical spin susceptibility.  In the 
following section, however, this statement is shown to be valid only  for 
the case of the short-range interaction, and  a completely 
different behavior shows up for the long-range interaction model.

%%%%%%%%%%%%%%%%%%%%%%%%%%%%%%%%%%%%%%%%%%%%%%%%%%%%%%%%%%%%%%%%%%%%%%%%%%%%
%%%%%%%%%%%%%%%%%%%%%%%%%%%%%%%%%%%%%%%%%%%%%%%%%%%%%%%%%%%%%%%%%%%

%%%%%%%%%%%%%%%%%%%%%%%%%%%%%%%%%%%%%%%%%%%%%%%%%%%%%%%%%%%%%%%%%%%%%%%%%%%%
%%%%%%%%%%%%%%%%%%%%%%%%%%%%%%%%%%%%%%%%%%%%%%%%%%%%%%%%%%%%%%%%%%%

%%%%%%%%%%%%%%%% TABLE %%%%%%%%%%%%%%%%
%\begin{table}
%\begin{center}
%\begin{tabular}{c|cccc} \hline
% & \makebox[8mm]{2$\Delta_{c}^{-}$} & \makebox[8mm]{$2\Delta_{s}^{-}$} 
% & \makebox[8mm]{2$\Delta_{o}^{-}$} & \makebox[8mm]{$2\Delta_{so}^{-}$}
%\\ \hline
%single orbital & 
%$\frac{1}{4}$ & $\frac{1}{2}$ & - & - \\ \hline
%two orbital ($\Delta$=0)&
%$\frac{1}{16}$ & $\frac{3}{4}$ & - & - \\ \hline
%two orbital($\Delta$$\ne$0) & 
%$\frac{1}{16}$ & $\frac{1}{4}$ & 0.218 & $\frac{1}{4}$
% \\
%\hline
%\end{tabular}
%\end{center}
%\caption{Critical exponent of the dynamical spin susceptibility; 
%2$\Delta_{\nu}^{-}$}
%\label{tbl:exponent2}
%\end{table}
%%%%%%%%%%%%%%%%%%%%%%%%%%%%%%%%%%%%%%%%%%%%%%%%%%%%%%%%%%%%%%%%%%%

%%%%%%%%%%%%%%%%%%%%%%%%%%%%%%%%%%%%%%%%%%%%%%%%%%%%%%%%%%%%%%%%%%%
%%%%%%%%%%%%%%%%%%%%%%%%%%%%%%%%%%%%%%%%%%%%%%%%%%%%%%%%%%%%%%%%%%%
%%%%%%%%%                   　　　　     %%%%%%%%%%%%%%%%%%%%%%%%%%
%%%%%%%%% 3. The long-range interacting  %%%%%%%%%%%%%%%%%%%%%%%%%%
%%%%%%%%%    case         　　　　       %%%%%%%%%%%%%%%%%%%%%%%%%%
%%%%%%%%%                   　　　　     %%%%%%%%%%%%%%%%%%%%%%%%%%
%%%%%%%%%%%%%%%%%%%%%%%%%%%%%%%%%%%%%%%%%%%%%%%%%%%%%%%%%%%%%%%%%%%
%%%%%%%%%%%%%%%%%%%%%%%%%%%%%%%%%%%%%%%%%%%%%%%%%%%%%%%%%%%%%%%%%%%

\section{Multicomponent Wigner Crystal}

Let us now discuss the model with the long-range interaction,
which may be important for
quantum wires as well as carbon nanotubes.
%\cite{Egger,Kane,Yoshioka,Iijima,Saito,Nakanishi} 
Also, for 1D highly correlated metallic systems close to
the Mott insulator, the screening effect of the bare 
interaction may be rather poor, so that the long-range
interaction is expected to play an important role. 
We will mainly focus on the latter case in the 
following discussions.

It is known that  physical properties
 in the presence of the long-range interaction are quite 
different from those of the TL liquid. For example, the charge density 
correlation function has a power-law decay in 
the TL liquid with the 
dominant contribution of the $2k_{F}$ oscillating 
piece. When the long-range interaction 
is present, the correlation decays 
much more slowly, and thus causes a dominant $4k_{F}$ oscillating
correlation.  This is sometimes referred to as a 1D 
version of the Wigner crystal\cite{Schulz}. 

We introduce the following interaction 
in  the Hamiltonian (\ref{HB_EQ}),
%%%%%%%%%%%%%%%%%%%%%%%%%%%%%%%%%%%%%%%%%%%%%%%%%%%%%%%%%%%%%%%%%%%%%%%%%
\begin{eqnarray}
U(x) = \frac{e^{2}}{\epsilon \sqrt{x^{2}+d^{2}}}
        {\rm exp} \left(
                    -\frac{\sqrt{x^{2}+d^{2}}}{\lambda}
                  \right),
\label{int_eq}
\end{eqnarray}
%%%%%%%%%%%%%%%%%%%%%%%%%%%%%%%%%%%%%%%%%%%%%%%%%%%%%%%%%%%%%%%%%%%%%%%%%%
which smoothly interpolates the short-range interaction 
($\lambda$$\rightarrow$$ 0$)
and the $1/x$-type long-range interaction
($\lambda$$\rightarrow$$\infty$)\cite{Otani}.
Here  the  cut-off of $x$$\sim$$d$ is introduced 
to avoid a singular behavior at short distance. 
The Fourier transformation of (\ref{int_eq}) gives the 
modified Bessel function $K_{0}(x)$, 
%%%%%%%%%%%%%%%%%%%%%%%%%%%%%%%%%%%%%%%%%%%%%%%%%%%%%%%%%%%%%%%%%%%%%%%%%%
\begin{eqnarray}
U(p) = \frac{2 e^{2}}{\varepsilon}
        K_{0} \left(
               \sqrt{p^{2}+{\lambda^{'}}^{-2}}
              \right),
\label{F-int_eq}
\end{eqnarray}
%%%%%%%%%%%%%%%%%%%%%%%%%%%%%%%%%%%%%%%%%%%%%%%%%%%%%%%%%%%%%%%%%%%%%%%%%%
where $\lambda^{'}$=$\lambda/d$. 

Even when the long-range Coulomb interaction is present, we 
can still diagonalize the Hamiltonian as far as
the Umklapp and backward scatterings are irrelevant,
leading to the formula (\ref{L-TL_eq}).
Note that $\tilde{K}_{c}(p)$ and $\tilde{K}_{o}(p)$  now
depend on the momentum $p$, whereas
$\tilde{K}_{s}(p)$ and $\tilde{K}_{so}(p)$ are independent of $p$, 
whose value is restricted to $\tilde{K}_{s}(p)$=$\tilde{K}_{so}(p)$=1
as before.

To discuss the effect of the long-range interaction 
on the system with the
orbital degeneracy, we again calculate the single-particle spectrum 
and the dynamical spin susceptibility. 
The single-particle Green function has the form, 
%%%%%%%%%%%%%%%%%%%%%%%%%%%%%%%%%%%%%%%%%%%%%%%%%%%%%%%%%%%%%%%%%%%%%%%%%%%%%%%
\begin{eqnarray}
 && G_{r a \sigma}^{<}(x,t) 
   \sim 
   \frac{1}{2 \pi}
    \prod_{\mu=c,o}
    {\rm e}^{-{\rm i} C_{\mu}(x,t)} {\it e}^{- D_{\mu}(x,t)}
\nonumber \\
  && \times
    \prod_{\nu=s,so}
\frac{{\rm e}^{-{\rm i} \frac{\pi}{2} 2 \tilde{\Delta}_{\nu}^{+} 
                                    {\rm sgn}(x+\tilde{{\it v}}_{\nu}t)}
      {\rm e}^{{\rm i} \frac{\pi}{2} 2 \tilde{\Delta}_{\nu}^{-} 
                                    {\rm sgn}(x-\tilde{{\it v}}_{\nu}t)}
        \left(
            \frac{\pi}{\beta \tilde{{\it v}}_{\nu}}
        \right)^{2 x_{\nu}}}
           { \left|
              \sinh     
              \left[       
                    \frac{\pi}{\beta \tilde{{\it v}}_{\nu}}
                    (x+\tilde{{\it v}}_{\nu}t)
              \right]
             \right|^{2\tilde{\Delta}_{\nu}^{+}}
             \left|
              \sinh     
              \left[       
                    \frac{\pi}{\beta \tilde{{\it v}}_{\nu}}
                    (x-\tilde{{\it v}}_{\nu}t)
              \right]
             \right|^{ 2\tilde{\Delta}_{\nu}^{-} } },
\nonumber \\
\label{LR-GF}
\end{eqnarray}
%%%%%%%%%%%%%%%%%%%%%%%%%%%%%%%%%%%%%%%%%%%%%%%%%%%%%%%%%%%%%%%%%%%%%%%%%%%%%%%
where
%%%%%%%%%%%%%%%%%%%%%%%%%%%%%%%%%%%%%%%%%%%%%%%%%%%%%%%%%%%%%%%%%%%%%%%%%%%%%%%
\begin{eqnarray}
C_{\mu}(x,t) 
&=& \int_{0}^{\infty} dp \frac{{\rm e}^{-\alpha' p}}{p} 
          \{
            2 \tilde{\Delta}_{\mu}^{+}(p) \sin [
                     \left\{x + \tilde{{\it v}}_{\mu}(p) t \right\}p ]
%\nonumber \\
%      && \ \ \ \ \ \ \ \ \ \ \ \ \ \ \ \ 
         +  2 \tilde{\Delta}_{\mu}^{-}(p) \sin [
                     \left\{x - \tilde{{\it v}}_{\mu}(p) t \right\}p ]
           \},
\nonumber \\
\\
D_{\mu}(x,t) &=& \int_{0}^{\infty} dp \frac{{\rm e}^{-\alpha' p}}{p} 
           (1 + 2 n_{p})
\nonumber \\
      && \ \ 
       \times   \{
            2 \tilde{\Delta}_{\mu}^{+}(p) \left(1 - \cos [
                     \left\{x + \tilde{{\it v}}_{\mu}(p) t \right\}p ]
                                  \right)
%\nonumber \\
%      && \ \ \ \ \ \ 
         +  2 \tilde{\Delta}_{\mu}^{-}(p) \left(1 - \cos [
                     \left\{x - \tilde{{\it v}}_{\mu}(p) t \right\}p ]
                                  \right)
           \},
\nonumber \\
\label{eq_CD}
\end{eqnarray}
%%%%%%%%%%%%%%%%%%%%%%%%%%%%%%%%%%%%%%%%%%%%%%%%%%%%%%%%%%%%%%%%%%%%%%%%%%
with  $\alpha'$=$a_{0}/d$ and 
$n_{p}$=1/[exp$\{$$\tilde{{\it v}}_{\mu}(p)$$p$/$k_{B}$T$\}$-1]. 
In contrast to the short-range case, 
we cannot integrate the 
charge as well as the orbital parts analytically,
so that we evaluate these integrals  numerically. 

\begin{figure}[thb]
\begin{center}
\vspace{0.5cm}
%\hspace{0.0cm}
\leavevmode \epsfxsize=150mm 
\epsffile{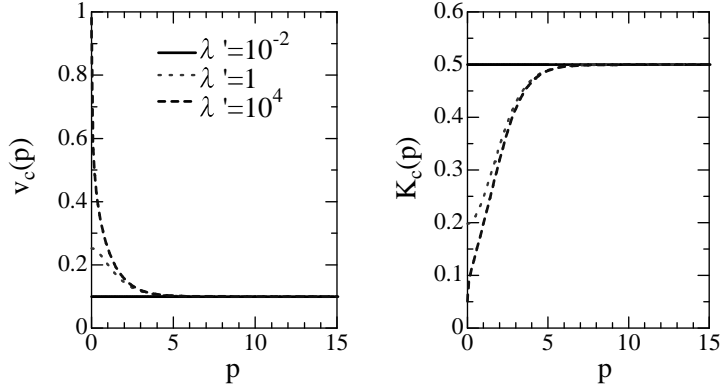}
%\vspace{-3.5cm}
\end{center}
\caption{Charge velocity ${\it v}_{c}(p)$ and TL parameter $K_{c}(p)$ 
for the single-orbital model with the long-range interaction. }
\label{2comLR-vckc}
\end{figure}

We investigate three cases for which the range of the interaction is
 long ($\lambda^{'}$=$10^{4}$), intermediate ($\lambda^{'}$=$1$),
and short  ($\lambda^{'}$=$10^{-2}$).
As the long-range interaction has no essential effects on 
 the spin velocity so it is given by that of the short-range case. 
We show the momentum-dependence of 
the charge velocity ${\it v}_{c}(p)$ and the TL  parameter $K_{c}(p)$
 calculated 
%${\it v}_{c}(p)$=${\it v}_{F}$$[1+2U(p)/\pi{\it v}_{F}]^{1/2}$, 
%$K_{c}(p)$=$[1+2U(p)/\pi{\it v}_{F}]^{-1/2}$ 
for the single-orbital case are shown in Fig. $\ref{2comLR-vckc}$, which
are scaled  so as to properly reproduce 
$\tilde{{\it v}}_{c}$=0.1 and $\tilde{K}_{c}$=1/2 in the case of
the short-range interaction.  It is seen  that 
when the long-range interaction is present, 
they are strongly dependent on $p$ in the small $p$ region. 
%%%%%%%%%%%%%%%%%%%%%%%%%%%%%%%%%%%%%%%%%%%%%%%%%%%%%%%%%%%%%%%%%%%%%%%%%%%%
%%%%%%%%%%%%%%%%%%%%%%%%%%%%%%%%%%%%%%%%%%%%%%%%%%%%%%%%%%%%%%%%%%%%%%%%%%%%
%%%%%%%%%%%%%%%%%%%%%%%%%%%%%%%%%%%%%%%%%%%%%%%%%%%%%%%%%%%%%%%%%%%%%%%%%%%%
\begin{figure}[thb]
\begin{center}
%\vspace{2.0cm}
%\hspace{0.0cm}
\leavevmode \epsfxsize=150mm 
\epsffile{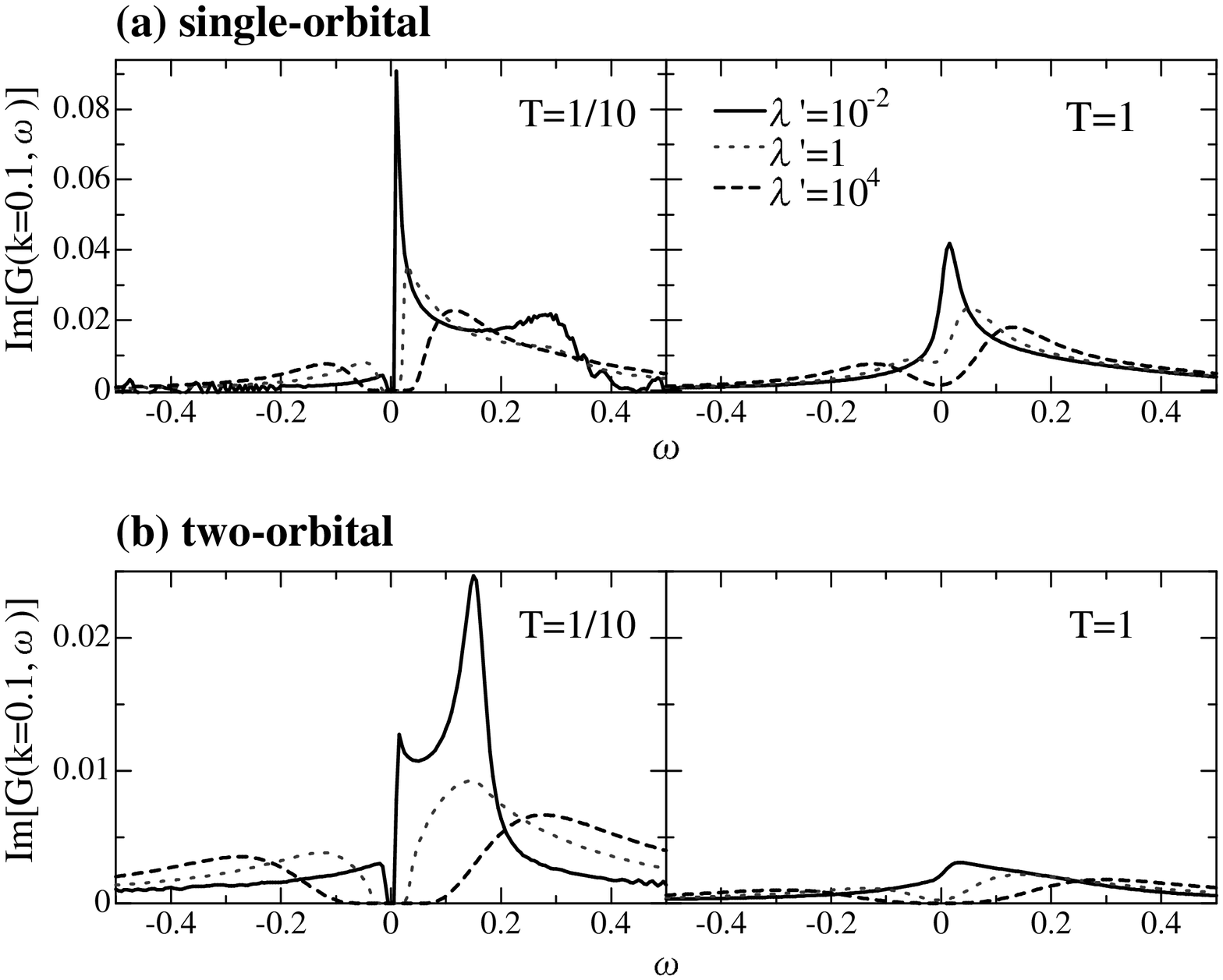}
%\vspace{-0.5cm}
\end{center}
\caption{Single-particle spectrum for 
(a) single-orbital model (${\it v}_{s}$=$\pi$) and 
(b) two-orbital model (${\it v}_{s}$=${\it v}_{o}$=${\it v}_{so}$=$\pi$/2). 
The range of the interaction is changed systematically 
by modifying $\lambda'$.}
\label{2+4comgfLR}
\end{figure}
%%%%%%%%%%%%%%%%%%%%%%%%%%%%%%%%%%%%%%%%%%%%%%%%%%%%%%%%%%%%%%%%%%%%%%%%%%%%

\begin{figure}[thb]
\begin{center}
%\vspace{-0.cm}
%\hspace{0.0cm}
\leavevmode \epsfxsize=150mm 
\epsffile{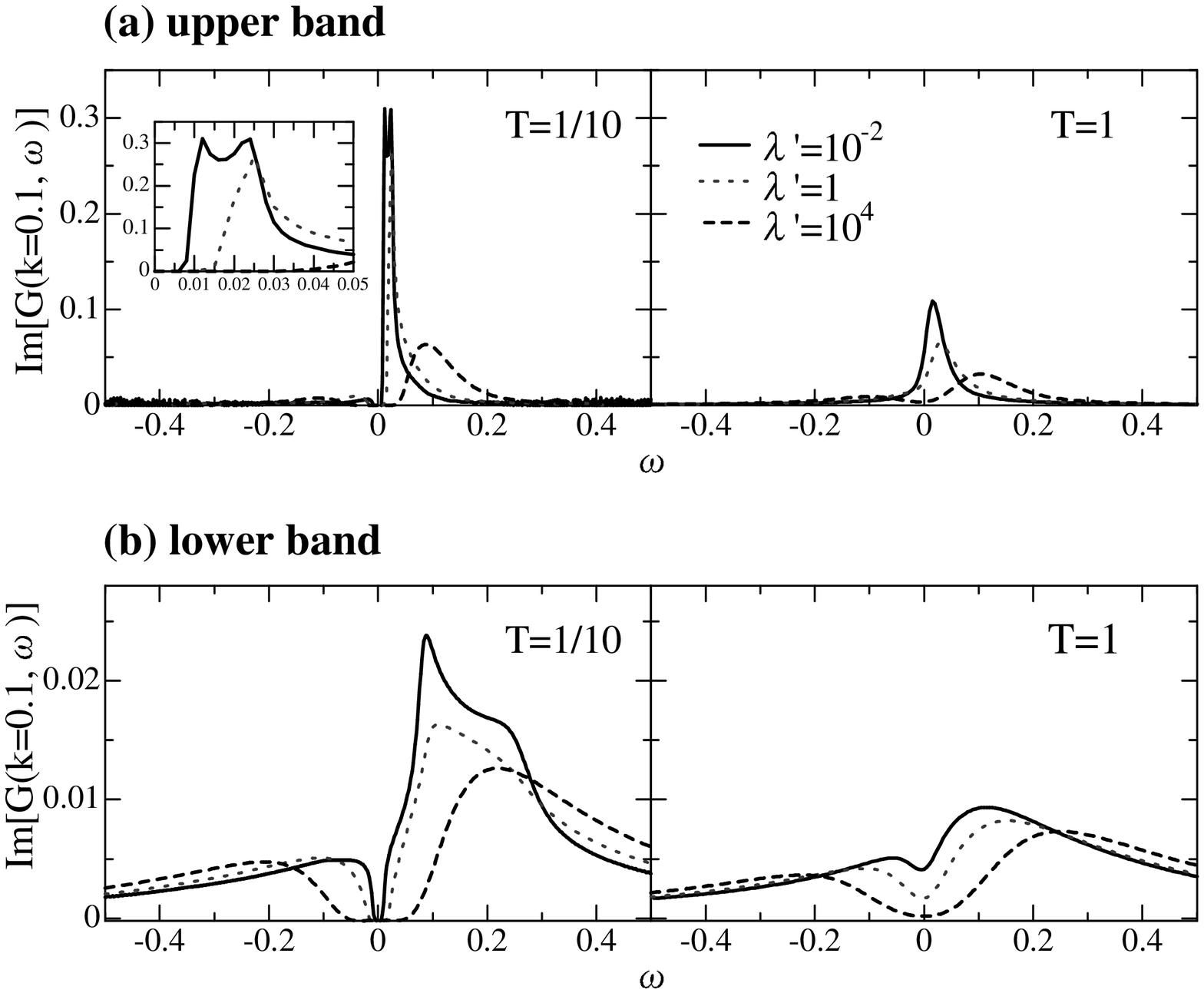}
%\vspace{-0.5cm}
\end{center}
\caption{Single-particle spectrum 
for the model with short as well as the long-range interaction.
The band splitting $\Delta$$\simeq$0.6, 
$\tilde{{\it v}}_{s}$$\simeq$2.6, $\tilde{{\it v}}_{so}$$\simeq$0.25.
}
\label{4comgfLR+o}
\end{figure}

The single-particle spectrum for  the single- as well as
two-orbital models is shown in Fig. $\ref{2+4comgfLR}$, 
 from which we can see
how the spectrum changes its profile in the 
presence of the long-range interaction. 
In the single-orbital case shown in  Fig. $\ref{2+4comgfLR}$(a),
both peaks coming from  the charge  ($\omega$$\sim$0.01) 
and spin sectors ($\omega$$\sim$0.3) are gradually suppressed
as the interaction  becomes long-ranged  (large $\lambda^{'}$).
This means that  the long-range interaction 
smoothly drives the system from the TL liquid to 
the 1D Wigner crystal.  In Fig. $\ref{2+4comgfLR}$(b), 
we  show  the two-orbital case without the
band splitting ($\Delta$=0).  Although
the spectral function  is affected considerably  
by the orbital degeneracy in the 
short-range interaction,  the 
spectrum loses such characteristic features
as the interaction becomes long-ranged.
Such smearing effect of the spectral function is also
seen for the case with the band splitting.
The single-particle spectrum of 
the lower and upper bands is shown 
 for $\Delta$$\simeq$0.6 in Fig. $\ref{4comgfLR+o}$.
Therefore, as long as the single-particle spectrum is
concerned, we would conclude that the long-range interaction
obscures not only the charge mode but also the other spin and
orbital modes in the 1D quantum liquids.
However, this naive
conclusion is misleading, which should not correctly capture
the effect of the long-range interaction.

%%%%%%%%%%%%%%%%%%%%%%%%%%%%%%%%%%%%%%%%%%%%%%%%%%%%%%%%%%%%%%%%%%%%%%%%%%%%
%%%%%%%%%%%%%%%%%%%%%%%%%%%%%%%%%%%%%%%%%%%%%%%%%%%%%%%%%%%%%%%%%%%%%%%%%%%%

\begin{figure}[thb]
\begin{center}
%\vspace{-0.0cm}
%\hspace{0.0cm}
\leavevmode \epsfxsize=150mm 
\epsffile{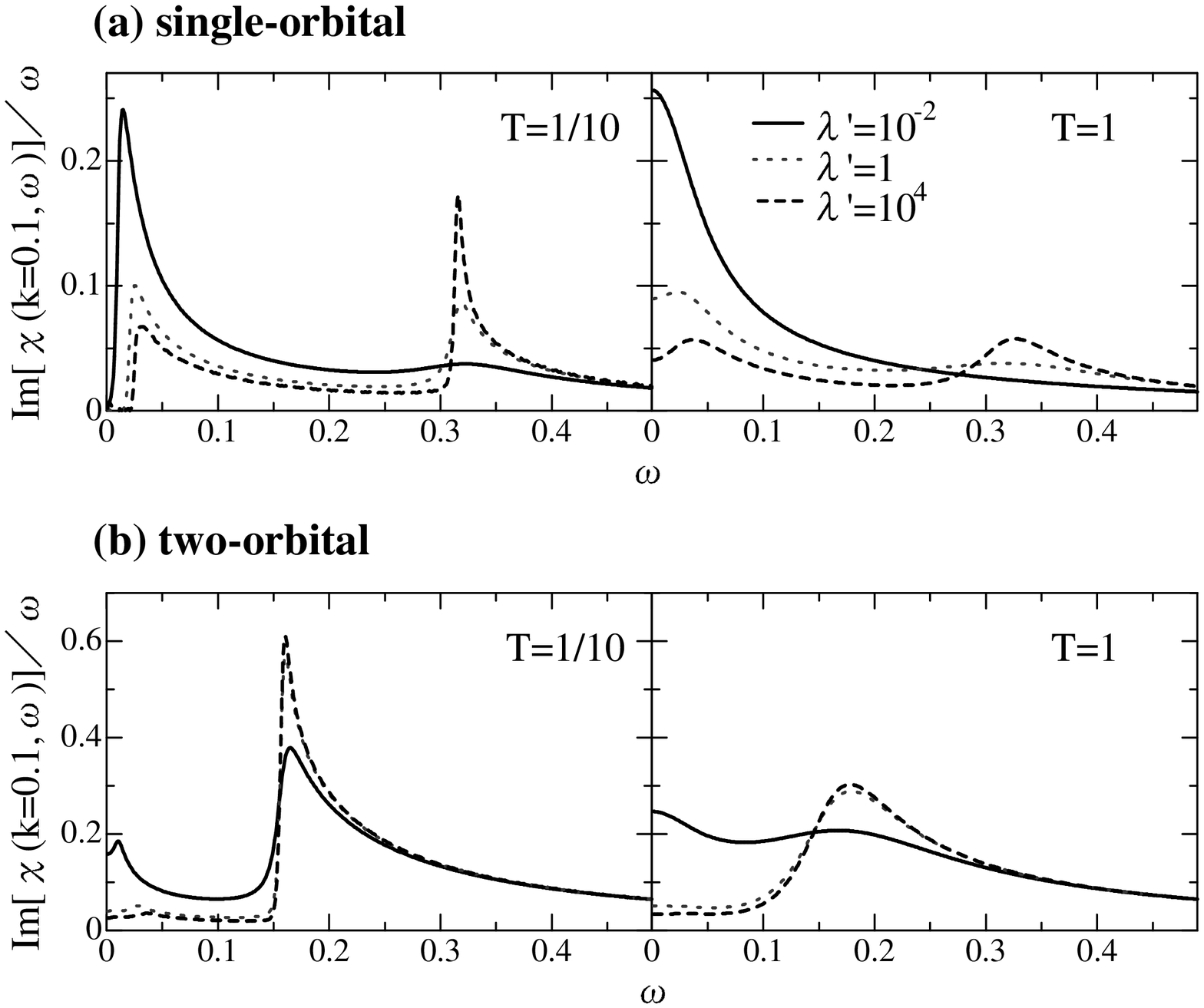}
%\vspace{-0.5cm}
\end{center}
\caption{Spin susceptibility with the short as well as the long-range 
interaction. }
\label{2+4comxiLR}
\end{figure}

To clarify the above point, let us 
consider the dynamical spin susceptibility.
In Fig. $\ref{2+4comxiLR}$, we show how the long-range 
interaction affects the dynamical spin
susceptibility with and without the orbital degeneracy.
As is the case for the single-particle spectrum, the 
peak structure for the charge sector is suppressed as
the interaction becomes long-ranged. However, 
it should be noticed that the peak structure due to the 
internal spin and orbital degrees freedom is {\it enhanced} 
as the long-range interaction is introduced,  as clearly
seen in Fig. $\ref{2+4comxiLR}$. This characteristic behavior  is quite 
contrasted to the results of the single-particle spectrum
discussed above.
The origin of the enhancement is rather clear; i.e. the long-range
interaction drives the system to the 1D Wigner crystal, which
almost freezes the charge degrees of freedom and 
thereby enhances the spin fluctuations effectively.
When the single-particle spectrum is considered, such enhanced
spin fluctuations are again masked by the charge excitations
which are accompanied by an electron-like excitation.

Therefore,  we come to the conclusion that the internal
 spin  and orbital fluctuations are indeed  enhanced by the
long-range interaction via the formation of the Wigner
crystal, although such enhancement is  
obscured in  the single-particle spectrum.  

%%%%%%%%%%%%%%%%%%%%%%%%%%%%%%%%%%%%%%%%%%%%%%%%%%%%%%%%%%
%%%%%%%%%%%%%%%%%%%%%%%%%%%%%%%%%%%%%%%%%%%%%%%%%%%%%%%%%%%%%%%%%%%

%%%%%%%%%%%%%%%%%%%%%%%%%%%%%%%%%%%%%%%%%%%%%%%%%%%%%%%%%%%%%%%%%%%%%%%%%%
%%%%%%%%%%%%%%%%%%%                          %%%%%%%%%%%%%%%%%%%%%%%%%%%%%
%%%%%%%%%%%%%%%%%%%    4. conclusion         %%%%%%%%%%%%%%%%%%%%%%%%%%%%%
%%%%%%%%%%%%%%%%%%%                          %%%%%%%%%%%%%%%%%%%%%%%%%%%%%
%%%%%%%%%%%%%%%%%%%%%%%%%%%%%%%%%%%%%%%%%%%%%%%%%%%%%%%%%%%%%%%%%%%%%%%%%%

\section{Summary}

We have investigated the effect of the orbital degeneracy 
and the band splitting on  the low-energy dynamical properties of
 1D quantum liquids in a metallic phase.
The single-particle spectrum and the dynamical spin susceptibility
have been calculated  by exploiting the bosonization
techniques. For the multicomponent TL liquid with the
short-range interaction, we have found that the spectral 
function is affected considerably by the orbital degeneracy
as well as the band splitting,  leading to characteristic 
features of the profile which are quite contrasted to
those expected for the Fermi liquid.
The effects of the long-range interaction have also been discussed. 
It has been found that the charge fluctuations are suppressed
while the internal spin and orbital fluctuations are enhanced 
by the long-range 
interaction via the formation of the Wigner crystal.
However, the enhanced spin fluctuations are hindered in 
 the single-particle spectrum by the almost frozen 
charge degrees of freedom.
It may be interesting to experimentally check whether 
such enhanced spin fluctuations can be indeed 
observed for 1D correlated electron systems,
such as the carbon nanotube, etc, 
for which the long-range interaction plays an
important role.

%%%%%%%%%%%%%%%%%%%%%%%%%%%%%%%%%%
\section*{Acknowledgements}
%%%%%%%%%%%%%%%%%%%%%%%%%%%%%%%%%%%%%
This work was partly supported by a Grant-in-Aid from the Ministry 
of Education, Science, Sports and Culture of Japan. 
A part of computations was done at the Supercomputer Center at the 
Institute for Solid State Physics, University of Tokyo
and Yukawa Institute Computer Facility. 
A. K.  was supported by 
Japan Society for the Promotion of Science. 

%%%%%%%%%%%%%%%%%%%%%%%%%%%%%%%%%%%%%%%%%%%%%%%
\section*{Appendix}
%%%%%%%%%%%%%%%%%%%%%%%%%%%%%%%%%

The $p$-dependence of $\alpha(p)$ and $y(p)$ is obtained as, 
\begin{eqnarray}
&& \tan [2\alpha(p)]=
  \frac{2 \Delta v}
  {v_1}(y(p)-y(p)^{-1})^{-1},   \nonumber \\
&& y^2=
   \frac{K_c(p)^{-2}+1}{K_o^{-2}+1},
\nonumber
\end{eqnarray}
%%%%%%%%%%%%%%%%%%%%%%%%%%%%%%%%%%%%%%
and the TL parameters, $K_c(p)$, $v_c(p)$, $K_o$ and $v_o$,
are given by  
%%%%%%%%%%%%%%%%%%%%%%%%%%%%%%%%%%%%%%%%%%%%%%%%%%%%%%%%%%%%%%%%%%%%%%%%%%
\begin{eqnarray}
K_{c}(p) &=& \left[
           1 + 4 U(p) a_{0} / \pi {\it v}_{1} 
             - V a_{0} / \pi {\it v}_{1}
          \right]^{-1/2}, 
\nonumber \\
K_{o} &=& \left[
           1 - V a_{0} / \pi {\it v}_{1}
          \right]^{-1/2},  
\nonumber \\
{\it v}_{c}(p) &=& {\it v}_{1} 
          \left[
           1 + 4 U(p) a_{0} / \pi {\it v}_{1} 
             - V a_{0} / \pi {\it v}_{1}
          \right]^{1/2}, 
\nonumber \\
{\it v}_{o} &=& {\it v}_{1}
          \left[
           1 - V a_{0} / \pi {\it v}_{1}
          \right]^{1/2}, 
\nonumber
\label{TLP_eq}
\end{eqnarray}
%%%%%%%%%%%%%%%%%%%%%%%%%%%%%%%%%%%%%%%%%%%%%%%%%%%%%%%%%%%%%%%%%%%%%%%%%%
where we have introduced 
 ${\it v}_{1}$=[${\it v}_{F}^{(1)}$+${\it v}_{F}^{(2)}$]/2 
and  
$\Delta{\it v}$=[${\it v}_{F}^{(1)}$-${\it v}_{F}^{(2)}$]/2.
Here $V$ is the on-site Coulomb interaction. 

Furthermore, the parameters $\tilde{{\it v}}_{\nu}(p)$ and 
$\tilde{K}_{\nu} (p)$ ($\nu$=$c$,$o$) 
 in the diagonalized Hamiltonian (\ref{L-TL_eq})
are given by, 
%%%%%%%%%%%%%%%%%%%%%%%%%%%%%%%%%%%%%%%%%%%%%%%%%%%%%%%%%%%%%%%%%%%%%%%%%%
\begin{eqnarray}
%\lefteqn{\tilde{{\it v}}_{c}^{2}(p)} 
%&&\nonumber \\
\tilde{{\it v}}_{c}^{2}(p)
&=& 
              \left\{
                \frac{{\it v}_{c}(p)}{K_{c}(p)} \xi_1(p)^2
              + \frac{{\it v}_{o}}{K_{o}} \xi_2(p)^2
              + 2 \Delta {\it v} \xi_1(p) \xi_2(p)
              \right\}
\nonumber \\
&& \times  \left\{
                {\it v}_{c}(p) K_{c}(p) \xi_1(p)^2 
              + {\it v}_{o} K_{o} \xi_3(p)^2 
              + 2 \Delta {\it v} \xi_1(p)\xi_3(p)
              \right\},
\nonumber 
\\
%\lefteqn{\tilde{{\it v}}_{o}^{2}(p)} &&
%\nonumber \\
\tilde{{\it v}}_{o}^{2}(p) 
&=& 
              \left\{
                \frac{{\it v}_{c}(p)}{K_{c}(p)} \xi_3(p)^2 
              + \frac{{\it v}_{o}}{K_{o}} \xi_1(p)^2
              - 2 \Delta {\it v} \xi_1(p)\xi_3(p)
              \right\}
\nonumber \\
&& \times  \left\{
                {\it v}_{c}(p) K_{c}(p) \xi_2(p)^2
              + {\it v}_{o} K_{o} \xi_1(p)^2
              - 2 \Delta {\it v} \xi_1(p)\xi_2(p)
              \right\},
\nonumber 
\\
%\lefteqn{\tilde{K}_{c}^{2}(p)} &&
%\nonumber \\
\tilde{K}_{c}^{2}(p) 
&=& 
\frac{
%              \left\{
                {\it v}_{c}(p) K_{c}(p) \xi_1(p)^2 
              + {\it v}_{o} K_{o} \xi_3(p)^2 
              + 2 \Delta {\it v} \xi_1(p)\xi_3(p)
%              \right\}
}
{
%\nonumber \\
%&& /          \left\{
                \frac{{\it v}_{c}(p)}{K_{c}(p)} \xi_1(p)^2 
              + \frac{{\it v}_{o}}{K_{o}} \xi_2(p)^2  
              + 2 \Delta {\it v} \xi_1(p) \xi_2(p)
%              \right\}
},
\nonumber 
\\
%\lefteqn{\tilde{K}_{o}^{2}(p)} &&
%\nonumber \\
\tilde{K}_{o}^{2}(p) 
&=& 
\frac{
%              \left\{
                {\it v}_{c}(p) K_{c}(p) \xi_2(p)^2 
              + {\it v}_{o} K_{o} \xi_1(p)^2 
              - 2 \Delta {\it v} \xi_1(p) \xi_2(p)
%              \right\}
}
{
%\nonumber \\
%&& /          \left\{
                \frac{{\it v}_{c}(p)}{K_{c}(p)} \xi_3(p)^2 
              + \frac{{\it v}_{o}}{K_{o}} \xi_1(p)^2 
              - 2 \Delta {\it v} \xi_1(p) \xi_3(p)
%              \right\}
}.
\nonumber
\end{eqnarray}
\end{document}